\documentclass[%
 reprint,
 amsmath,amssymb,
 aps,showkeys,https://www.overleaf.com/project/608bd335d666745eb22d6f68
]{revtex4-2}

\usepackage{xcolor}

\usepackage{siunitx}
\usepackage{graphicx}
\usepackage{dcolumn}
\usepackage{bm}
\usepackage{hyperref}
\usepackage{multirow}
\begin{document}

\title{Bayesian parameter-estimation of Galactic binaries in LISA data with Gaussian Process Regression}

\author{Stefan H. Strub}
 \email{stefan.strub@erdw.ethz.ch}
\author{Luigi Ferraioli}%
\author{Cedric Schmelzbach}%
\author{Simon C. Stähler}%
\author{Domenico Giardini}%
\affiliation{%
Institute of Geophysics, ETH Zurich\\ Sonneggstrasse 5, 8092 Zurich, Switzerland
}%


\begin{abstract}
The Laser Interferometer Space Antenna (LISA), which is currently under construction, is designed to measure gravitational wave signals in the milli-Hertz frequency band. It is expected that tens of millions of Galactic binaries will be the dominant sources of observed gravitational waves. The Galactic binaries producing signals at mHz frequency range emit quasi monochromatic gravitational waves, which will be constantly measured by LISA. To resolve as many Galactic binaries as possible is a central challenge of the upcoming LISA data set analysis. Although it is estimated that tens of thousands of these overlapping gravitational wave signals are resolvable, and the rest blurs into a galactic foreground noise; extracting tens of thousands of signals using Bayesian approaches is still computationally expensive. We developed a new end-to-end pipeline using Gaussian Process Regression to model the log-likelihood function in order to rapidly compute Bayesian posterior distributions. Using the pipeline we are able to solve the Lisa Data Challenge (LDC) 1-3 consisting of noisy data as well as additional challenges with overlapping signals and particularly faint signals.
\end{abstract}

\keywords{Gravitational Waves, LISA, Galactic Binaries, Gaussian Process Regression}
\maketitle


\section{Introduction}
Gravitational wave (GW) science is a relatively young field, marked by the first observation of a GW in 2015 by the LIGO detector \cite{abbott2019gwtc}. The field is growing with multiple new detectors under construction or in planning state aiming for increased sensitivity and different frequency ranges. Current observations with ground based detectors, LIGO \cite{aasi2015advanced} and Virgo \cite{acernese2014advanced}, are limited to high frequency bands of $[10,10^4]$ Hz due to seismic noise and their limited size \cite{Martynov_2016}. The Laser Interferometer Space Antenna (LISA), currently under construction, is a space based interferometric system with arm lengths of 2.5 million km, capable of detecting lower frequency GWs in the $[0.1, 100] \, \mathrm{mHz}$ range that is not contaminated by terrestrial seismic and anthropogenic noise sources \cite{amaro2017laser}. While existing terrestrial sensors were restricted to detecting individual events, like white dwarf mergers, with signals above 10 Hz, LISA will unlock the galactic background hum, which will require new source extraction methods \cite{nelemans2001gravitational}.

The dominant sources in the LISA frequency band will be tens of millions of Galactic binaries (GB) emitting quasi monochromatic gravitational waves. Since the sources are far away from merger, their gravitational waves will be constantly measured during the operational time of LISA. It is estimated that tens of thousands of these overlapping signals are resolvable, while the rest blurs into a galactic foreground noise. The challenge of this analysis will be to extract as many sources as possible, where estimating the parameters of GBs accurately provides valuable information for studying the dynamical evolution of binaries \cite{taam1980gravitational, willems2008probing, nelemans2010chemical, littenberg2019binary, piro2019inferring}.

A variety of methods have been proposed for the extraction of GB signals. One group of methods is based on maximum likelihood estimation (MLE) to find the best match to the data as investigated by \cite{zhang2021resolving} using swarm optimization and by \cite{bouffanais2016DE} employing a hybrid swarm based and differential evolution algorithm. Another group of methods uses the Bayesian approach where the posterior distribution describes the uncertainty of the source parameters. The most successful Bayesian approaches are Markov chain Monte Carlo (MCMC) based methods, such as blocked annealed Metropolis-Hastings (BAM) \cite{PhysRevD.75.043008, crowder2007genetic, littenberg2011detection} an MCMC algorithm with simulated annealing, and recently a pipeline with a time-evolving source catalog as new data from LISA arrives \cite{PhysRevD.101.123021}. The posterior distribution provides additional information but is computationally more expensive to obtain compared to searching for the MLE.

Here, we present a novel way to combine the best of both, MLE and Bayesian approaches, by solving the GB search problem in two steps. Firstly, we use an optimization algorithm to search for the global optimum by exploiting the speed of MLE methods. In the second step, we apply the Metropolis-Hastings algorithm to sample the posterior distribution around the MLE to estimate the uncertainty of the obtained solution. Metropolis-Hastings algorithms, require the expensive computation of the log-likelihood for multiple parameter sets. The FastGB algorithm provided by the LDC \cite{LDC} based on \cite{cornish2007tests}, is already a fast LISA response approximation but is still computationally expensive if millions of evaluations are needed to compute the posterior distribution of one signal. To speed up the sampling process, we use Gaussian Process Regression (GPR) to model the log-likelihood function using the FastGB response to avoid simulating the LISA response for each sample.

In Section \ref{sec:bayes} we introduce the general description of Bayesian parameter estimation used for LISA data. A detailed description of our new end-to-end pipeline to calculate a posterior distribution from a simulated LISA data stream is provided in Section \ref{sec:pipeline}. Then, we present the results of the pipeline tackling the LISA Data Challenge (LDC)1-3 and other challenging signal extraction tasks in Section \ref{sec:results}. Finally, in Section \ref{sec:conclusion} we discuss the performance of the current pipeline and potential extensions of the pipeline.

\section{Bayesian signal extraction for LISA data}
\label{sec:bayes}
\begin{figure}[!ht]
\includegraphics[width=0.485\textwidth]{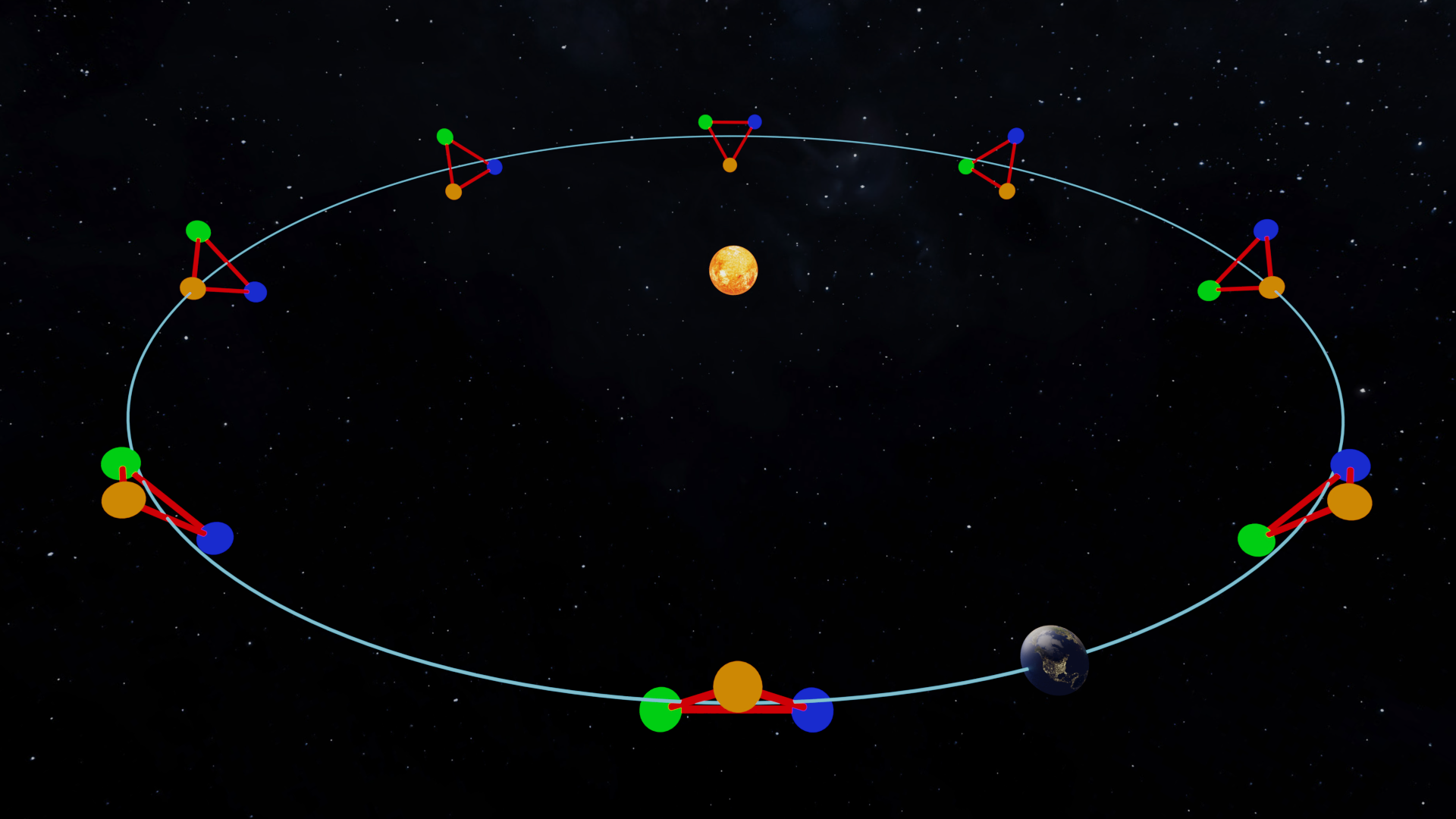}
\caption{\label{fig:LISA} During one year, the three space crafts of LISA, shown as green, orange, and blue dots, orbit the sun while conducting one full cartwheel-like motion. LISA's constellation is pictured at eight different points in time with 1.5 month intervals. The blue line marks the orbits of LISA and earth. The size of the orbit is shrunk by a factor of ten and the sizes of earth and LISA spacecrafts are not to scale, but the remaining distances between the objects and the size of the sun are at scale to each other. This picture and a visualization of LISA measuring an exaggerated GW are accessible at \cite{strub_stefan_2022_6761175}.}
\end{figure}

LISA will consist of a triangular arrangement of three satellites with arm lengths of $L= \SI{2.5}{million \, km}$. The satellites will be orbiting the sun trailing earth while completing one cartwheel-like motion during the course of one year as depicted in Figure \ref{fig:LISA}. Due to this motion, the recording of GWs will be frequency, amplitude and phase modulated \cite{cornish2003lisa}.

The measured LISA data $d(t)$ can be written as $d(t) = s(t, \theta) + n(t)$, where $s(t, \theta)$ is the gravitational wave signal and $n(t)$ is the noise. For simplicity we introduce the following notation $d = d(t)$ and $s(\theta) = s(t,\theta)$ dropping the time dependence. In order to extract the parameter set $\theta$ describing the signal $s(\theta)$ from noisy data we can use Bayes' Theorem

\begin{equation}
    p \left( \theta | d \right) = \frac{p \left( d | \theta \right) p \left( \theta \right)}{p \left(d \right)} 
\end{equation}
where the posterior distribution $  p \left( \theta | d \right)$ describes the conditional probability distribution of the parameters $\theta$ after measuring $d$. On the right hand side, $p \left( \theta \right)$ is called the prior distribution, where already known distributions of parameters can be included in the posterior distribution. The model evidence $p \left(d \right)$ is independent of $\theta$. Therefore, $p \left(d \right)$ does not influence the relative probability. For Markov chain Monte Carlo methods, which we will use in our pipeline, the evidence will be a normalization factor such that $\int p \left( \theta | d \right) \, d\theta = 1$. Finally, the last term $p \left( d | \theta \right)$, known as the likelihood, is the probability of measuring the data stream $d$ for given parameters $\theta$. In GW data analysis the log-likelihood is defined as

\begin{equation}
\label{eq:loglikelihood}
\log p(d \mid \theta) = -\frac{1}{2}  \langle  d-s\left(\theta \right) |  d-s\left(\theta \right) \rangle,
\end{equation}
where the scalar product is defined as

\begin{equation}
    \langle x \left( t \right) |y \left( t \right) \rangle = 4 \mathcal{R} \left( \int_0^\infty  \frac{\tilde{x}\left( f \right) \tilde{y}^*\left( f \right)}{S\left( f \right)} \, df \right),
\label{scalar product}
\end{equation}
where $\mathcal{R}$ denotes the real part of the argument, $S \left( f \right)$ is the one sided power spectral density of the noise \cite{PhysRevD.82.022002} and the Fourier transform is

\begin{equation}
    \tilde{x} \left( f \right) = \int_{-\infty}^\infty  x \left( t \right) e^{-i 2 \pi f t} dt .
\end{equation}

 To cancel the laser noise, the Laser measurements of the LISA arms will be combined using time-delay-interferometry (TDI) to three observables X, Y, Z \cite{tinto1999cancellation, Armstrong_1999, estabrook2000time, dhurandhar2002algebraic, tinto2014time}. Therefore, the data $d$ and signal $s(\theta)$ consist of TDI responses that have multiple channels and we adjust the inner product to
 \begin{equation}
 \langle  d-s\left(\theta \right) |  d-s\left(\theta \right) \rangle = \sum_{\alpha \in \mathcal{M}} \langle  d_\alpha-s_\alpha\left(\theta \right) |  d_\alpha-s_\alpha\left(\theta \right) \rangle 
 \end{equation}
 where $ \mathcal{M} = \left\{ X, Y, Z \right\}$ as the default TDI or $\mathcal{M} = \left\{ A, E, T \right\}$ with 
  \begin{equation}
    \begin{aligned}
 A &= \frac{1}{\sqrt{2}} \left( Z - X \right) \\
 E &= \frac{1}{\sqrt{6}} \left( X - 2Y + Z \right) \\
 T &= \frac{1}{\sqrt{3}} \left( X + Y + Z \right)
    \end{aligned}
 \end{equation}
which are noise uncorrelated observables \cite{vallisneri2005synthetic}. Ultimately we use $A,E,T$ to calculate the log-likelihood where we only consider $A,E$ for signals with frequencies $f < f_\ast = 1/\left(2 \pi L \right) \simeq \SI{19.1}{mHz}$ since the gravitational wave response for $T$ would be suppressed \cite{PhysRevD.101.123021}.
 
There are eight parameters $\theta = \left\{\mathcal{A}, \lambda, \beta, f, \dot{f}, \iota, \phi_0, \psi\right\}$ \cite{cornish2007tests} with which we simulate a GW from a GB, where $\mathcal{A}$ is the amplitude, the ecliptic longitude $\lambda$ and ecliptic latitude $\beta$ are the sky locations, $f$ is the frequency of the GW, $\dot{f}$ is the first order frequency derivative, $\iota$ is the inclination angle, $\phi_0$ is the initial phase and $\psi$ is the polarization angle. In this work, we neglect higher order frequency derivatives.

\section{The pipeline}
\label{sec:pipeline}
\begin{figure*}
\includegraphics[width=1\textwidth]{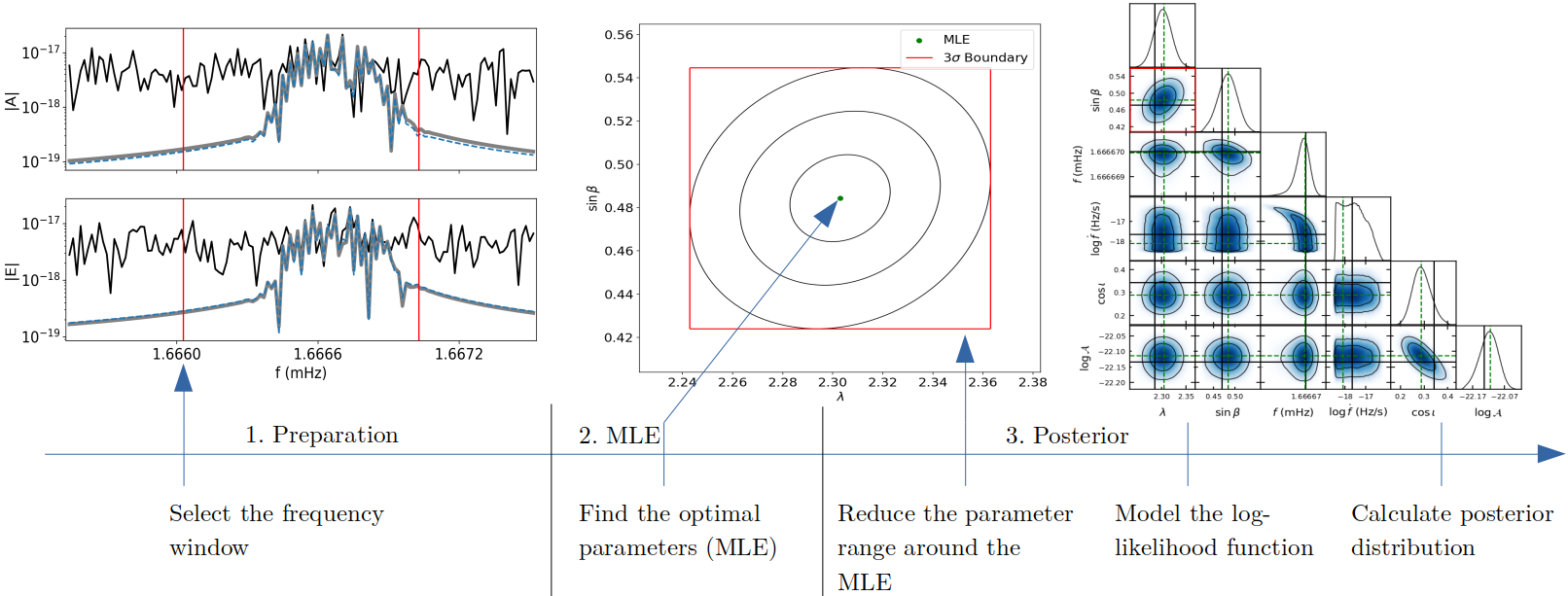}
\caption{\label{fig:pipeline} Graphical illustration of the GW detection and GB-parameter estimation pipeline starts on the left with the LISA data and progresses to the right with the posterior distribution of the parameters as the output. The function of each step is mentioned along the flow chart with the three pictures visualizing their outputs. The left picture shows two LISA strains $|A|$ and $|E|$ as black solid lines which are the input of the pipeline. To compare the end result, the plot also shows the true signal as a grey solid line which is mostly covered by the found signal shown as the blue dashed line. The first step of the pipeline selects a small window where the frequency range boundaries are shown as red vertical lines. Next, a global optimization algorithm searches for the parameters with the simulated signal matching the data best. This is shown in the middle plot as the green dot marking the maximum likelihood estimate of the sky localization parameters. Afterwards, we can estimate the $\SI{1,2,3} {\sigma}$ contour lines shown as black ellipses using the Fisher Information Matrix. Then the red box, framing the $\SI{3}{\sigma}$ covariance region, determines the reduced parameter space for the next steps. Within this reduced parameter space, we model the log-likelihood function using Gaussian Process Regression. Finally, the Metropolis-Hastings algorithm calculates the posterior distribution, shown in the right figure, with the use of the modeled log-likelihood function. A larger version of the posterior distribution is shown in Figure \ref{fig:LDC1-3 corner}.}
\end{figure*}

Our end-to-end pipeline has the goal to extract GBs from simulated LISA data providing posterior distributions for each found source. Figure \ref{fig:pipeline} gives an overview of each step starting on the left and progressing to the right. The pipeline consists of three main steps: part 1 is selecting a frequency window of the data that will be investigated for signal extraction and sets the parameter search range, part 2 is finding the maximum likelihood estimate (MLE), and part 3 is computing the posterior distribution around the identified MLE to give a more informative solution, which is the end goal of the signal extraction. Each step is described in the following subsections.

\subsection{Selecting a frequency window}
The first step of the pipeline is to select narrow frequency bands to analyze. A sliding window method is used to scan the whole frequency domain. The window width we use is $\SI{1}{\mu \mathrm{Hz}}$, which is large enough to include the expected signal of a single source, since GBs observed in the LISA band are quasi monochromatic. A frequency derivative of $\dot f = \SI{10^{-15}}{\mathrm{Hz}^2}$ observed during two years $T_{obs} = \SI{2}{years}$, results in a bandwidth of $\dot f \cdot T_{obs} \approx \SI{0.63}{\mu \mathrm{Hz}}$ which is small enough to be within the frequency window. Additionally, the GW frequencies are smeared due to Doppler shifts of $\Delta f \approx f \cdot \SI{10^{-4}}{\mathrm{Hz}}$ caused by the movement of LISA for an observation duration of one year or longer \cite{Vallisneri_2009}.

To ensure the proper detection of signals close to the boundaries we pad both sides by $\SI{0.5}{\mu \mathrm{Hz}}$ but only accept the signals which are detected to be within the unpadded window. For the signals within the LDC1-3 and tests covered in this paper it turned out to be sufficient to use the aforementioned constant window width and padding size, but the window sizes can be a function of frequency to adjust for wider and narrower signals.

\subsection{Global optimization to find the maximum likelihood estimate}

\label{section2}
A crucial part of the pipeline is the parameter optimization algorithm to search for the MLE $\theta_{max} = \arg \max_{\theta} \log p \left( d | \theta \right)$, where the log-likelihood function is defined in Equation \ref{eq:loglikelihood}. Rewriting the log-likelihood to 

\begin{equation}
\label{eq:log-likelihood p}
\log p(d \mid \theta) = -\frac{1}{2}  \langle  d |  d \rangle -\frac{1}{2}  \langle  s\left(\theta \right) |  s\left(\theta \right) \rangle + \langle  d |  s\left(\theta \right) \rangle,
\end{equation}
we see that we can neglect the term independent of $\theta$ and therefore might just as well maximize the log-likelihood ratio

\begin{equation}
\label{eq:log-likelihood ratio}
\log \mathcal{L} = -\frac{1}{2}  \langle  s\left(\theta \right) |  s\left(\theta \right) \rangle + \langle  d |  s\left(\theta \right) \rangle.
\end{equation}

Extracting the amplitude parameter from the signal $ s\left(\theta \right) = \mathcal{A} s\left(\theta_\mathcal{A} \right)$ where for $\theta_\mathcal{A} = \left\{\lambda, \beta, f, \dot{f}, \iota, \phi_0, \psi\right\}$ we obtain

\begin{equation}
\log \mathcal{L} = -\frac{1}{2} \mathcal{A}^2  \langle  s\left(\theta_\mathcal{A} \right) |  s\left(\theta_\mathcal{A} \right) \rangle + \mathcal{A} \langle  d |  s\left(\theta_\mathcal{A} \right) \rangle,
\end{equation}
which we can maximize over the amplitude $\mathcal{A}$ analytically

\begin{equation}
\frac{\partial \log \mathcal{L}}{\partial \mathcal{A}} = - \mathcal{A} \langle  s\left(\theta_\mathcal{A} \right) |  s\left(\theta_\mathcal{A} \right) \rangle + \langle  d |  s\left(\theta_\mathcal{A} \right) \rangle = 0,
\end{equation}

\begin{equation}
\mathcal{A}_{max} =  \frac{\langle  d |  s\left(\theta_\mathcal{A} \right) \rangle}{\langle  s\left(\theta_\mathcal{A} \right) |  s\left(\theta_\mathcal{A} \right) \rangle }.
\end{equation}

As a result, the search is used to maximize a quantity independent of $\mathcal{A}$

\begin{equation}
\log \mathcal{L}_{\max \mathcal{A}} = \frac{\langle  d |  s\left(\theta_\mathcal{A} \right) \rangle^2}{ 2 \langle  s\left(\theta_\mathcal{A} \right) |  s\left(\theta_\mathcal{A} \right) \rangle } = \frac{\rho^2}{2},
\end{equation}
where $\rho$ marks the signal-to-noise ratio (SNR)
\begin{equation}
\label{eq:SNR}
\rho =  \frac{\langle  d |  s\left(\theta \right) \rangle}{\sqrt{\langle  s\left(\theta \right) |  s\left(\theta \right) \rangle}} =  \frac{\langle  d |  s\left(\theta_\mathcal{A} \right) \rangle}{\sqrt{\langle  s\left(\theta_\mathcal{A} \right) |  s\left(\theta_\mathcal{A} \right) \rangle}}.
\end{equation}

With noisy data, there are almost always parameter sets that match the data better than the signal with the true parameters

\begin{equation}
    \mathcal{L}(\theta_{max}) \geq \mathcal{L}(\theta_{true}).
\end{equation}
Therefore, it is expected that $\theta_{max}$ provides a good estimate of $\theta_{true}$ but $\theta_{max} \neq \theta_{true}$. To better describe this mismatch, we calculate the probability distribution of the parameters in the third part of the pipeline.

Finding $\theta_{max}$ is a non-convex problem where methods such as gradient descent can likely get stuck in a local optimum. To find the global optima more time consuming search methods are needed.

We use the prior distributions shown in Table \ref{tab:prior}. All prior distributions are uniform except for $\beta$, $\iota$ and $\dot{f}$. The prior is uniform in the sine ecliptic latitude $\sin \beta$, and cosine inclination $\cos \iota$. The prior of the frequency derivative is log-uniform $\log \dot{f}$ and to determine the boundaries we use \cite{PhysRevD.101.123021}

\begin{equation}
    \dot{f} = \frac{96}{5} \pi^{8/3} \mathcal{M}^{5/3} f^{11/3}
\end{equation}

where $\mathcal{M}$ is the chirp mass and  $f$ the frequency.
For the upper boundary the masses of the binary are set to the Chandrasekhar limit $m_1 = m_2 = \SI{1.4}{M_\odot}$ and for the lower limit we set $\mathcal{M} = \SI{0.1}{M_\odot}$. The frequency prior is uniform in the frequency window determined by the first section of the pipeline with a width of $\Delta f = \SI{1}{\mu Hz}$. Since we determine the amplitude $\mathcal{A}$ analytically we do not require a prior.

\begin{table}[!ht]\footnotesize
\caption{Boundaries of the prior distribution.}
\begin{ruledtabular}
\begin{tabular}{ccc}
           Parameter &  Lower Bound & Upper Bound  \\ \hline
    $\sin \beta$  &     $-1$   &  1\\
  $\lambda$&      $-\pi$ & $\pi$\\
  $\dot{f}$ &  $1.26 \cdot 10^{-10} f^{11/3}$  & $1.02 \cdot 10^{-6} f^{11/3}$ \\
         $\cos \iota$&      $-1$  & 1\\
        $\phi_0$&          0  & 2$\pi$\\
         $\psi$&          0  & $\pi$ \\
\end{tabular}
\label{tab:prior}
\end{ruledtabular}
\end{table}

\subsubsection{Coordinate Descent}

We obtained successfully the MLE using a variant of a coordinate descent search \cite{wright2015coordinate} which starts at a random point in the parameter space and then performs sequentially random sampling along one coordinate axis or coordinate hyperplane jumping to the max log-likelihood if a better value is found. The algorithm tends to sometimes stay at a local maximum depending on the starting parameter set.

The search is rather quick and therefore it is feasible to run multiple searches with different initial parameters either sequentially on a single CPU core or in parallel on multiple cores. Unsuccessful searches can be pruned as soon as they stagnate for a certain amount of iterations to free up computational resources. There are other pruning methods such as Asynchronous Successive Halving \cite{li2020massively} or median pruning which can be used.

The hyperparameters we found to be suitable for GB signal extraction are 2 dimensional search planes, with 50 random trials on each iteration, maximally 100 iterations per search, and around 50 - 100 searches with random initial parameters per signal. One search is set to be pruned after no improvement of 16 iterations. Additionally, a median pruning method is added as well, where at iterations 30, 40, 50, 60, and 70 the current log-likelihood value is compared to the mean log-likelihood value at the same iteration of previous searches. The search is aborted if the intermediate value is worse than the intermediate value of previous searches at the same step. The five best results are then locally optimized by sequential least squares programming using the SciPy library. The best result is then selected as the optimum. 

\subsubsection{Differential Evolution}
Another method to find the MLE we investigated is differential evolution (DE), which is an evolutionary algorithm. Genetic algorithms have already been used for LISA data analyses, namely for GB signal extraction \cite{crowder2007genetic} and for massive black hole binary signal extraction \cite{petiteau2010search}. A hybrid form of the DE and particle swarm algorithms was used for a 7 parameter ($\dot f = 0$) GB search by \cite{bouffanais2016DE}.

The DE algorithm is described in \cite{storn1997differential}. The used implementation is the off-the-shelve algorithm part of the SciPy library. The values of the hyperparameters are listed in Table \ref{tab:DE}.

\begin{table}[!ht]\footnotesize
\caption{Values of the DE hyperparameters.}
\begin{ruledtabular}
\begin{tabular}{lr}
     Parameter &  Value \\ \hline
    Strategy  &     'best1exp'\\
    Population size &    8 \\
    Relative tolerance &  $10^{-8}$ \\
    Max-iteration   &     1000\\
    Recombination   &     0.75\\
    Mutation range  &     $(0.5,1)$\\
\end{tabular}
\label{tab:DE}
\end{ruledtabular}
\end{table}

Comparing the two methods, the coordinate descent algorithm is well suitable for parallelization where one search uses multiple CPU cores. Thus coordinate descent has seemingly an advantage over the differential evolution algorithm. Though this advantage is neglectable by parallelizing multiple searches, where each CPU core is used for one search window at a time. Later in Section \ref{section:ldc1-3} we compare the reliability and speed of the two methods by extracting the 10 signals of the LDC1-3 on 20 test runs.

\subsection{Determining the reduced parameter space for calculating the posterior distribution}
\label{FIM}
The posterior distribution lies typically in a small volume of the parameter space. Therefore, it is not needed to sample the log-likelihood beyond a certain boundary in order to accurately calculate the posterior. The goal here is to find the boundaries of the parameter space containing the main part of the posterior distribution. 

The Fisher Information Matrix (FIM) is a popular choice to estimate the parameter uncertainty for LISA signals. The FIM can be computed by 
\begin{equation}
    F_{ij} = \langle \partial_i s(\theta) | \partial_j s(\theta) \rangle
\end{equation}
without much computational cost where $\partial_i$ is the partial derivative with respect to the $i$\textsuperscript{th} component of the parameter $\theta$ and the scalar product is defined at Equation \ref{scalar product}. The matrix inversion of the FIM computed at the MLE gives the Gaussian covariance matrix of the log-likelihood function at the MLE. The assumption of a Gaussian distribution is not the case for all parameters and is therefore not a good approximation. Nonetheless, the approximation can be used to set the desired parameter boundaries within which we want to model the log-likelihood function. For computing the derivatives of the FIM we use the second order forward finite difference method with a step size of $10^{-9}$ times the search space determined by the prior.

For our goal, we set the volume such that it typically includes the $\SI{3}{\sigma}$ standard deviation, as shown as the red box in the middle plot of Figure \ref{fig:pipeline}. For the frequency, it is sufficient to set the boundary to the $\SI{1}{\sigma}$ standard deviation. Since the polarization and initial phase are degenerate we neglect their distribution and set the search space as a narrow range in order to reduce the complexity of modeling the log-likelihood. Therefore, we set $\psi \in [ \psi_{\mathrm{MLE}} - \frac{\pi}{1000},\psi_{\mathrm{MLE}} + \frac{\pi}{1000} ]$ and $\phi_0 \in [ \phi_{0,\mathrm{MLE}} - \frac{2\pi}{1000},\phi_{0,\mathrm{MLE}} + \frac{2\pi}{1000}]$.

Sampling the frequency derivative $\dot{f}$ in log scale creates the issues of large standard deviations $\sigma$ if $\dot{f}$ is small and therefore the $\SI{3}{\sigma}$ range would span the whole prior. To avoid this problem we set the reduced search region to be $\log \dot{f} \in [ -18.5, -16]$ if the lower bound of the reduced $\dot{f}$ region is below $-16$. This region has to be adjusted if the observation time differs from two years and is later motivated by Equation \ref{eq:threshold}.

Similarly if $\cos \iota \approx \pm 1$ the standard deviation of $\cos \iota$ and $\mathcal{A}$ become large and the resulting $\SI{3}{\sigma}$ range would span beyond their prior boundaries. Therefore if part of the reduced $\cos \iota$ region is $< -0.9$ or $> 0.9$ we set $\cos \iota \in [ - 1, - 0.7]$ respectively $\cos \iota \in [ 0.7, 1]$ and the amplitude to $\log \mathcal{A} \in [ \log \mathcal{A}_{\mathrm{MLE}} - 0.1 l_{\mathcal{A}}  , \log \mathcal{A}_\mathrm{MLE} + 0.1 l_{\mathcal{A}} ]$ where $l_{\mathcal{A}}$ is the width of the amplitude search space in log-scale. The amplitude boundaries are determined by the SNR $\rho \in [7,1000]$ which, according to \cite{PhysRevD.101.123021}, is related to the amplitude via

\begin{equation}
\mathcal{A} \left( \rho \right) = 2 \rho \left( \frac{T_{obs} \, \sin^2\left(f / f_\ast \right)}{S\left(f\right)} \right)^{-1/2}.
\end{equation}
Therefore we obtain $l_{\mathcal{A}} = \log \mathcal{A} \left( 1000 \right) - \log \mathcal{A} \left( 7 \right) $.

In general, if the MLE is close to the initial boundary, the resulting reduced region could go beyond the initial boundary. In that case, the reduced region would be cut at the initial boundaries in order to not exceed them.

The aforementioned ranges for the boundary reduction can be adjusted to be larger or smaller, where larger ranges require more computational power to accurately model and evaluate the log-likelihood function but give a larger coverage of the posterior distribution.

\subsection{Modelling the log-likelihood function}

The bottleneck to rapidly computing the posterior distribution is the calculation of the log-likelihood for each sample (see Equation \ref{eq:log-likelihood ratio}). Since for each parameter set the corresponding LISA response has to be computed and matched to the data, which is repeated $\mathcal{O} (10^6)$ times for a precise posterior distribution. Therefore it is advantageous to model the log-likelihood function and use the model's approximation to rapidly compute the posterior distribution.

Gaussian Process Regression (GPR) is a method to predict a continuous variable, $\log \mathcal{L}$ in our case, as a function of one or more dependent variables, where the prediction takes the form of a probability distribution.

Generally, a Gaussian Process is a collection of random variables $h(x)$, indexed by a set $x \in X$, where any finite number of which have a joint Gaussian distribution \cite{Rasmussen2006}. Any Gaussian Process is completely specified by the mean function $m : X \rightarrow \mathbb{R}$ and covariance function $k(x,x'): X \times X \rightarrow \mathbb{R}$. For any training set $X_t \subseteq X$, $X_t = \{x_1, ... , x_n\}$ it holds that $h(X_t) = \left[ h(x_1), ..., h(x_n)\right] \sim \mathcal{N} \left( \mathbf{m}(X_t) , K(X_t,X_t) \right)$ where

\begin{equation*}
\mathbf{m}(X_t) = \begin{bmatrix}
           m(x_{1}) \\
           m(x_{2}) \\
           \vdots \\
           m(x_{n})
         \end{bmatrix},
         \\
\end{equation*}

\begin{equation*}
K(X_t,X_t) = \begin{pmatrix}
k(x_1,x_1) &\hdots & k(x_1,x_n)\\
\vdots &\vdots &\vdots  \\
k(x_2,x_1) &\hdots & k(x_n,x_n)
\end{pmatrix}
\end{equation*}
and $\mathcal{N}\left(\mathbf{m},K \right)$ denotes the Gaussian distribution with mean vector $\mathbf{m}$ and covariance matrix $K$.

With noise-free observations $\mathbf{y} = h(X_t)$, we can predict the function values $h(X_P)$ at desired locations $X_p \subseteq X$ as they are related by Gaussian distributions

\begin{eqnarray*}
    h(X_p) \sim && \mathcal{N} \left( K(X_p,X_t)K(X_t,X_t)^{-1} \mathbf{y}, \right. \\ && K(X_p,X_p) -  \left. K(X_p,X_t)K(X_t,X_t)^{-1}K(X_t,X_p) \right).
\end{eqnarray*}

\begin{figure}
\includegraphics[width=0.5\textwidth]{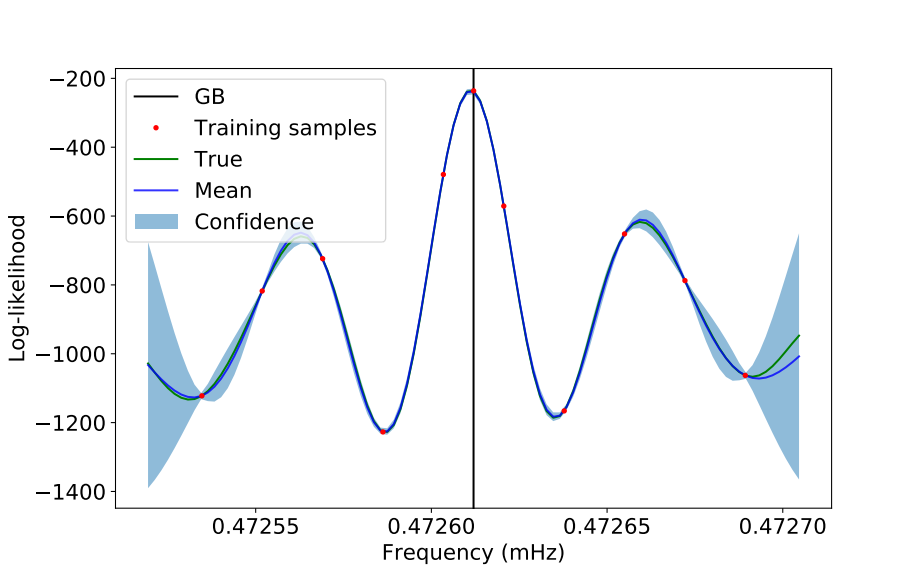}
\caption{\label{fig:gpr} Gaussian Process Regression is used to model the log-likelihood as a function of frequency while the remaining parameters are set constant as the true values. The black line indicates the true frequency. The target function is shown in green representing the true log-likelihood values. The model is trained with 11 samples shown as red dots. The resulting mean of the model is shown as the blue line and the $\SI{2}{\sigma}$ confidence as the blue shaded region.}
\end{figure}

In Figure \ref{fig:gpr} we show such a predictive distribution of the log-likelihood as a function of frequency, where the remaining parameters are constant. Using only 11 training samples located at frequencies $X_t$ with computed $\mathbf{y}= \log \mathcal{L}(X_t)$, shown as red dots, the model is able to accurately predict the target function. The prediction is displayed as a mean function with the shaded region marking the confidence interval. Note how the prediction is less confident close to the boundary since no neighboring training samples beyond the boundary are available. The reader is referred to \cite{Rasmussen2006} for a more detailed discussion of Gaussian Processes.

GPR has already been applied for LISA data, namely the interpolation and marginalization of waveform error in extreme-mass-ratio-inspiral parameter estimation for up to three parameters \cite{Gaussian_processes}. Here we present the successful modeling of the log-likelihood function on all eight parameters of GBs. The biggest factor for successfully modeling in eight dimensions is the reduction of the parameter space in the previous step \ref{FIM} of the pipeline. Since the posterior distribution spans only across a small region of the parameter space, it is possible to restrict the log-likelihood model to a small parameter space of interest.

The model is trained on a set of computed log-likelihood values of parameter sets sampled randomly within the reduced parameter range. The model is trained with 1000 training samples and then tested on 500 different samples by comparing the model's prediction to the true log-likelihood. If the root-mean-square error (RMSE) is larger than a set threshold $\mathrm{RMSE} > 0.6$, we dismiss the current model and increase the training set by another 1000 samples to compute a new model. This procedure is repeated until $\mathrm{RMSE} < 0.6$.

In this pipeline we found the squared exponential or radial basis function (RBF) covariance function \cite{Rasmussen2006} to be the most accurate since the likelihood function is a smooth function. Periodicities, which could have been an advantage for periodic kernels, like the sky location parameters are cut by the reduction of the parameter space and therefore not suitable. The models are trained and evaluated using the GaussianProcessRegressor package of scikit-learn \cite{scikit-learn} which is based on Algorithm 2.1 of \cite{Rasmussen2006}. For increased stability, the boundaries of the inputs are normalized to $[0,1]$. The hyperparameters are listed in Table \ref{tab:GPR}.

\begin{table}[!ht]\footnotesize
\caption{Hyperparameters of the RBF kernel.}
\begin{ruledtabular}
\begin{tabular}{ccc}
           Parameter &  Initial length scale & Length scale bounds  \\ \hline
  $\log \mathcal{A}$&      1 & $\left[ 0.1,10 \right]$\\
    $\sin \beta$  &     2 & $\left[ 0.1,10 \right]$\\
  $\lambda$&     5 & $\left[ 0.1,10 \right]$\\
  $f$ &     1 & $\left[ 0.1,10 \right]$\\
  $\log \dot{f}$ &     1 & $\left[ 0.1,10 \right]$\\
         $\cos \iota$&     1 & $\left[ 0.1,10 \right]$\\
        $\phi_0$&      1 & $\left[ 0.1,30 \right]$\\
         $\psi$&      1 & $\left[ 0.1,30 \right]$\\
\end{tabular}
\label{tab:GPR}
\end{ruledtabular}
\end{table}

\subsection{Posterior distribution}

The last step in the pipeline is the computation of the posterior distribution. The choice here is the Metropolis-Hastings Monte Carlo (MHMC) algorithm, which proposes new parameters $\theta'$ according to a proposal distribution $g(\theta')$ which in general depends on the previous parameters $\theta$, therefore we write $g(\theta' \mid \theta)$ . The proposed parameters are then accepted with probability 
\begin{equation}
P(\theta',\theta)=\min \left(1, \left[ {\frac {p(d \mid \theta')}{p(d \mid \theta)}}{\frac {g(\theta\mid \theta')}{g(\theta'\mid \theta)}} \right] ^{\frac{1}{\mathcal{T}}} \right) 
\end{equation}
where $p(d \mid \theta)$ is the likelihood defined in Equation \ref{eq:log-likelihood ratio} and $\mathcal{T}$ is the temperature for simulated annealing.

Due to the curse of dimensionality, few samples would be accepted by sampling uniformly within the parameter space making the algorithm unfeasible for fast computation. Keeping in mind that the best proposal distribution is the posterior itself we use high temperature posteriors as the new proposal distribution. A first run of the MHMC at a high temperature with a uniform prior provides a first flattened posterior distribution, which then is used as the proposal distribution for a second run at a lower temperature. This procedure is repeated until the temperature $\mathcal{T}=1$ is reached, which yields the desired posterior distribution. 
We used four steps. The first three runs are with $10^5$ samples where the first run is at a temperature $\mathcal{T}=10$, the second at $\mathcal{T}=2$, the third one at $\mathcal{T}=1$ and then a final run with $\mathcal{T}=1$ and an increased number of samples $10^6$. The temperature lowering can also be continuously where the sampling at $\mathcal{T}>1$ is neglected for the final posterior distribution during the so called burn-in phase.

The posterior distribution for some parameters is not Gaussian distributed and therefore not suitable to set the proposal distribution as a Gaussian approximation of the posterior. Therefore it is advantageous to use a multivariate kernel density estimation (KDE) \cite{o2016fast, o2014reducing}. An eight dimensional KDE with millions of data points is intractable but also not necessary for our proposal distribution. There are only two parameter pairs $f$-$\dot f$ and $\mathcal{A}$-$\iota$ which have non-Gaussian distributions. Therefore it is enough to calculate the KDE of these pairs and multiply the probability of all pairs to get the final proposal distribution. The sky-locations are Gaussian distributed but since the non-Gaussian KDE for two parameters is computationally cheap, we use non-Gaussian KDE as well. Due to the degeneracy of the polarization and initial phase, they are fixed within a small search space without any influence on the search, therefore they can be sampled uniformly. The parameter pairs are listed in table \ref{tab:parameter-pairs}.

\begin{table}[!ht]\footnotesize
\caption{Parameter pairs for the proposal distribution.}
\begin{ruledtabular}
\begin{tabular}{ccccc}
    Parameter pair       & Distribution  \\ \hline
   $\mathcal{A}$-$\iota$ &         KDE  \\
   $f$-$\dot f$ &          KDE   \\
   $\lambda$-$\beta$ &          KDE \\
   $\phi_0$-$\psi$  &          uniform \\
\end{tabular}
\label{tab:parameter-pairs}
\end{ruledtabular}
\end{table}

\section{Results}
\label{sec:results}

The following results are obtained with the pipeline described above. To simulate the LISA response of Galactic binaries we used FastGB since the GB signals of the LDC1-3 \cite{LDC} are also simulated using FastGB. All presented simulations have an observation time of 2 years. The code of this pipeline and the following results are open source and available under the MIT license \cite{strub_code}.

\subsection{LDC1-3: Noisy non-overlapping signal search}
\label{section:ldc1-3}
First we evaluated the pipeline on the LDC1-3v2 \cite{LDC} data set which consists of simulated instrument noise and 10 non-overlapping GB signals. Comparing the two optimization methods in Table \ref{tab:search success} we find DE to have a higher success rate while the averaged number of evaluations are roughly the same. Therefore we use DE as the global optimization algorithm for further evaluations in this article.

\begin{table}[!ht]\footnotesize
\caption{Averaged number of function evaluations $n_{\textrm{fe}}$ required for each signal of LDC1-3. The signals are labeled as their true frequency. Each minimization is executed 20 times for each signal with randomly chosen initial parameters. A search is determined as successful if the likelihood of the found parameters is higher than the likelihood of the true parameters.}
\begin{ruledtabular}
\begin{tabular}{ccccc}
    Signal       & \multicolumn{2}{c}{Coordinate Descent}  & \multicolumn{2}{c}{Differential Evolution}  \\ \hline
 Frequency (mHZ) &  Success rate &   $n_{\textrm{fe}}$ &  Success rate &  $n_{\textrm{fe}}$ \\ \hline
   1.35962 &          0.75 & 24800 &           0.70 & 18881 \\
   1.25313 &          0.65 & 26175 &           0.90 & 19362 \\
   1.81324 &          0.70 & 23775 &           0.95 & 20029 \\
   1.66667 &          0.35 & 25325 &           0.85 & 22105 \\
   1.94414 &          0.30 & 27075 &           0.95 & 22995 \\
   3.22061 &          0.20 & 26900 &           1.00 & 23190 \\
   3.51250 &          0.35 & 27625 &           0.75 & 28411 \\
   1.68350 &          0.60 & 24550 &           0.95 & 22764 \\
   6.22028 &          0.05 & 27175 &           0.95 & 32407 \\
   2.61301 &          0.40 & 26925 &           1.00 & 21301 \\
\end{tabular}
\label{tab:search success}
\end{ruledtabular}
\end{table}

\begin{figure}
\includegraphics[width=0.49\textwidth]{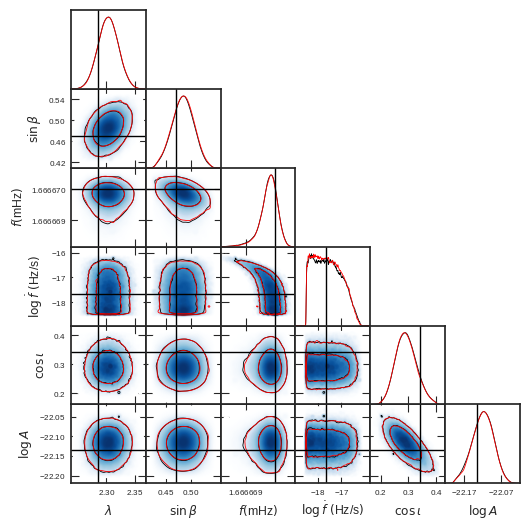}
\caption{\label{fig:LDC1-3 corner} Posterior distribution of a signal of the LDC1-3. The black lines show the true parameters. The blue contours enclose the 68\% and 95\% probability regions of the parameter estimation. The red contour lines are the posterior calculated using FastGB instead of GPR.}
\end{figure}

Next we show the computational speed up using the GPR for different training set sizes in Table \ref{tab:evaluation time}. The GPR prediction time increases with the size of the training set due to increased model size. Most models have a training set size of 1000 or 2000 when the prediction is accurate enough (RMSE $< 0.6$). Therefore compared to using FastGB a speed up of factor 2000 or 750 is most common and reduces the computational time significantly. In Figure \ref{fig:LDC1-3 corner} we compare the posterior distributions where for the blue distribution the log-likelihood is approximated using GPR with a training set size of 1000 and an RMSE of 0.36, and the red distribution is computed using FastGB. The two distributions pass the Gelman-Rubin convergence test \cite{gelman1992inference} with a threshold set to 1.003. The strong overlap confirms the accuracy of the log-likelihood approximation.

\begin{table}[!ht]\footnotesize
\caption{GPR log-likelihood evaluation time compared to the log-likelihood computation time using FastGB. The evaluation time is the average time required to predict the log-likelihood for one given parameter set using the GPR. The average is computed by taking the average of predicting $10^5$ random log-likelihoods for each of the LDC1-3 signals. The speed up decreases with the training set size $N_{train}$ and compares the evaluation time with the direct log-likelihood computation time of $\SI{30}{ms}$.}
\begin{ruledtabular}
\begin{tabular}{cccc}
 $N_{train}$ &  Training time (s) &  Evaluation time ($\mu $s) &   Speed up\\ \hline
   1000 &          4 &         15 & 2000  \\
   2000 &          19 &         40 & 750 \\
   3000 &          65 &         74 & 405 \\
   4000 &          107 &         111 & 270 \\
\end{tabular}
\label{tab:evaluation time}
\end{ruledtabular}
\end{table}

The global optimization time for the 10 signals was 8 min if it was known that there is only one signal per search window, otherwise, the search duration was 16 min. The posterior calculation took 20 min for 10 posteriors with 1 million samples each. The calculations are performed on a consumer grade laptop equipped with an Intel Core i9-9880H CPU.

\begin{figure}[!ht]
\includegraphics[width=0.49\textwidth]{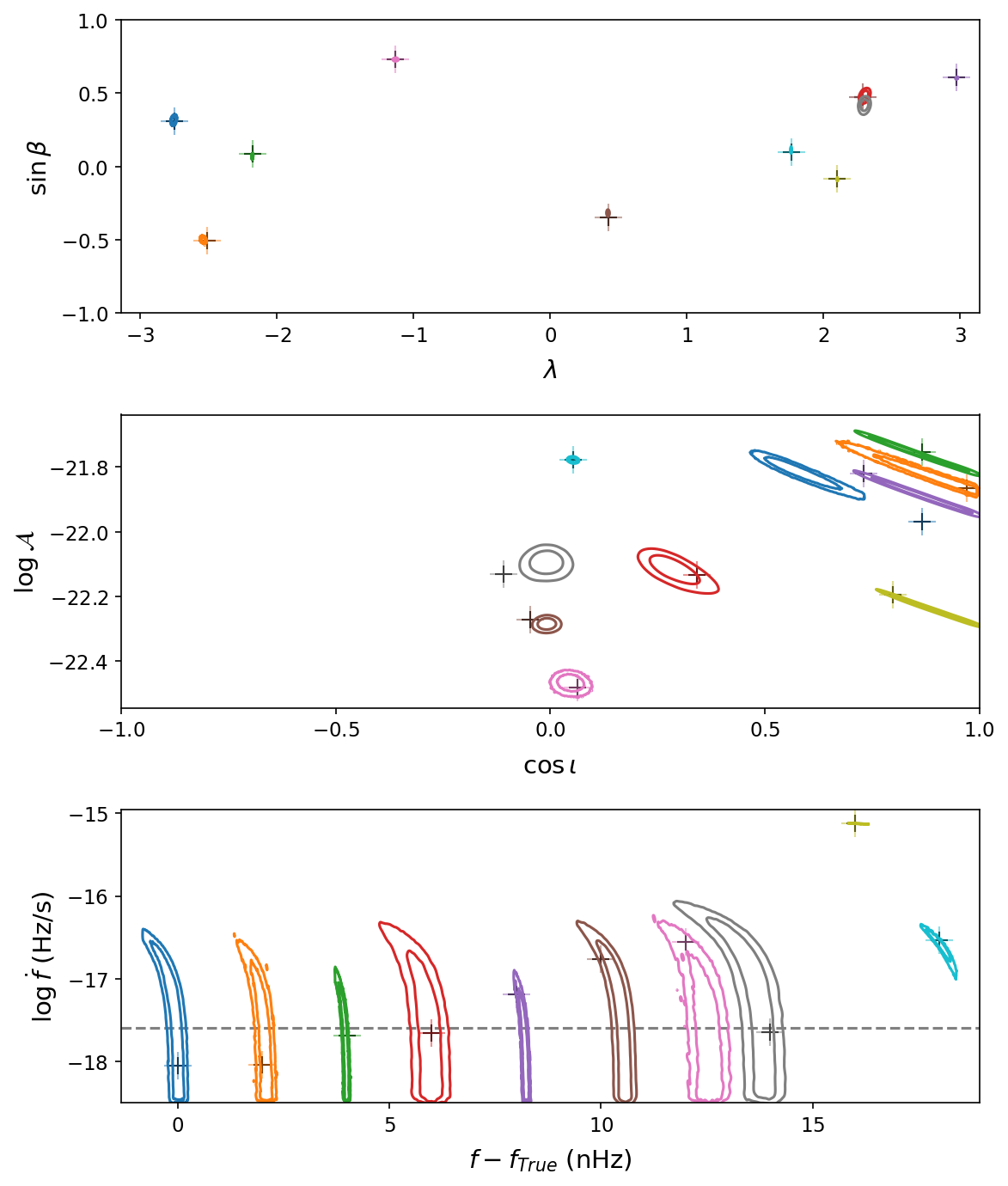}
\caption{\label{fig:global solution} Posterior distributions of LDC1-3. The black crosses represent the true values. They are shaded in the background of the cross matching the color of the found posterior distributions. Each color represents one signal. The contour lines enclose the 68$\%$ and 95$\%$ posterior probability regions. The dashed horizontal line in the bottom plot is at $\log ( 0.01 / T_{obs}^2) = -17.6$ for an observation time of $T_{obs} = 2 $ years. The frequencies of the 10 signals differ by $\approx \SI{0.01}{Hz}$ which makes the visualization of posterior widths of $\approx \SI{1}{nHz}$ unfeasible. Therefore the true frequency values are subtracted from the posterior and are frequency shifted by steps of $\SI{2}{nHz}$ to not overlap with each other. The initial phase and polarization are not shown since the solution is degenerate and therefore does not necessarily match the true values.}
\end{figure}

We show the true parameters and found posterior distributions for all 10 signals in Figure \ref{fig:global solution}. The initial phase $\phi_0$ and polarization $\psi$ are omitted since the degeneracy does not allow for unique solutions.
The amplitudes of the 10 signals are all in the same order of magnitude, where the remaining parameters are distributed across a wide range of the expected parameter ranges. As seen in the top plot of Figure \ref{fig:global solution}, the sky location is resolved for all signals with high precision. The middle plot shows the correlation between the amplitude and the inclination. Low inclination, which is high $\cos \iota$ in the plot, results in a non-Gaussian correlation.

The bottom plot shows the correlation of frequency $f$ and frequency derivative $\dot{f}$. Note, we show the logarithm of the frequency derivative $\log \dot{f}$ to visualize the effect of small frequency derivatives $\dot{f}$. The $\log$ scale broadens the visualized posterior distribution for small values even though $\dot{f}$ is resolved precisely.

Using the $\log$ scale we find that below the dashed line, the frequency derivative does not correlate with the frequency anymore. This suggests that a frequency derivative smaller than this threshold has no effect on the signal. If the total change of frequency during the observation time $T_{obs}$ is significantly smaller than the resolution in the frequency domain  $\dot{f} T_{obs}  \ll \Delta f_{res} = \frac{1}{T_{obs}}$, then the frequency deviation $\dot{f}$ has no measurable effect on the signal

\begin{equation}
    \dot{f} \ll \frac{1}{T_{obs}^2}.
\end{equation}

From the bottom plot, we can now estimate the threshold where the posterior distribution of $\log \dot{f}$ and $f$ is not curved anymore. The grey dashed line gives such an estimate, where the frequency derivative
 
\begin{equation}
\label{eq:threshold}
    \dot{f}  	\lessapprox \frac{0.01}{T_{obs}^2}
\end{equation}
 is not measurable anymore.
 
Therefore we can set a lower boundary $\log \dot{f} > \log ( 0.01 / T_{obs}^2) = -17.6$ for $T_{obs} = \SI{2}{years}$ for the posterior computation, since a larger search space would only increase the complexity of modelling the log-likelihood function and provides no additional information. Nonetheless, in this article we chose $\log \dot{f} > -18.5$ to better visualize the correlation of $\dot{f}$ and $f$.

The smooth distributions are due to the high sample count of $10^6$ samples. As a result of the high amount of samples, each posterior passed the Gelman-Rubin convergence test with a threshold of 1.003 and with four different chains.

An additional source of error is the approximation of the model used to simulate the LISA response such as FastGB in our chase. This error could be presented as additional uncertainty in the posterior distribution. Since with real data, the estimate of the approximation error is a challenge on its own we do not add this uncertainty.

\subsection{Extracting a faint signal}

To further evaluate the effectiveness of the pipeline, we further test the extraction of a faint signal. Here we use the SNR as defined in Equation \ref{eq:SNR} as a quantity of faintness, where a faint signal has a low SNR. To accept a found signal in a data set with instrument noise we determined an SNR $\rho > 8$ as a reliable acceptance rate. For fainter signals $\theta_{max}$ and $\theta_{true}$ deviate too much to be confident.

\begin{table}[!ht]\footnotesize
\caption{True parameters of a faint source and the parameters of the found MLE where the uncertainties are deducted from the posterior distribution shown in Figure \ref{fig:low snr}. The uncertainties of the initial phase and polarization are omitted due to the degeneracy. The SNR is shown for the true parameters and the MLE, where a higher SNR means the signal matches the data better.}
\begin{ruledtabular}
\begin{tabular}{ccc}
           Parameter &  Signal & MLE  \\ \hline
            $\mathcal{A} \times 10^{23}$  &  4.55 & $5.5^{+1.2}_{-1.6}$ \\
    $\beta$  &     0.2   &  $0.17 \pm 0.15$\\
    $\lambda$&      1.5 & $1.5 \pm 0.07$\\
            $f$ (mHz) &  1.40457  & $1.4045697^{+1.9\times 10^{-6}}_{-1.5\times 10^{-6}}$\\
  $\log \dot{f}$ &  -17.3  & $-16.2^{+0.3}_{-2.3}$ \\
         $\iota$&      1.2  & $1.33^{+0.1}_{-0.18}$\\
         $\phi_0$&          3  & $3.06$\\
         $\psi$&          2  & $1.875$\\
     SNR  & 7.93  &  8.25 \\
\end{tabular}
\label{tab:parameters low snr}
\end{ruledtabular}
\end{table}

\begin{figure}[!ht]
\includegraphics[width=0.49\textwidth]{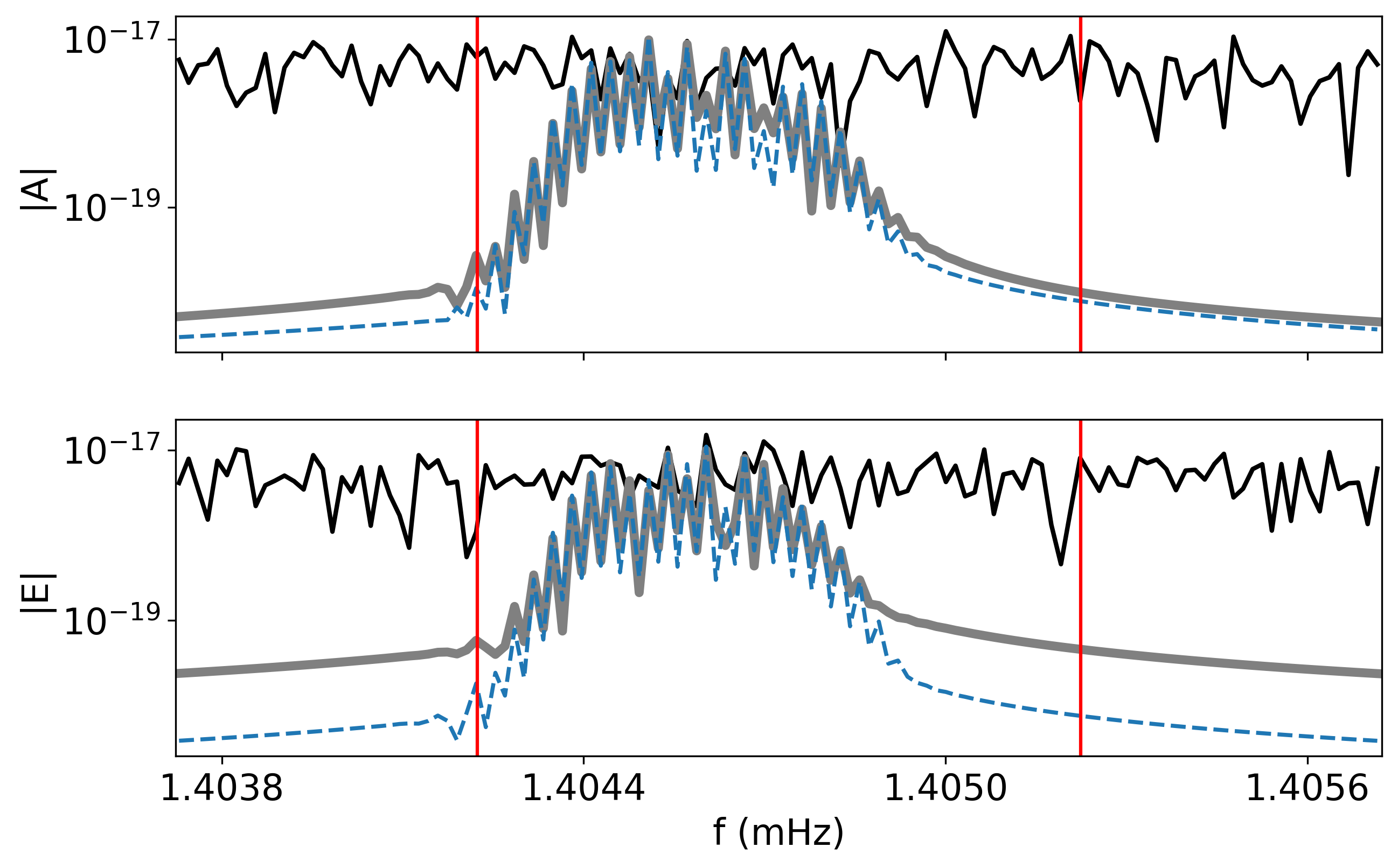}
\caption{\label{fig:low snr signal and mle}  The data consists of a noisy part of the LDC1-3v2 and a low SNR signal. The data is shown as the black solid line and the signal within the data is the grey solid line. The pipeline finds the MLE shown as the dashed blue line. The parameters of the true and found signal are listed in Table \ref{tab:parameters low snr}. The red vertical lines are the frequency boundaries with an interval of 1 $\mu \mathrm{Hz}$ in between.}
\end{figure}

\begin{figure}[!ht]
\includegraphics[width=0.49\textwidth]{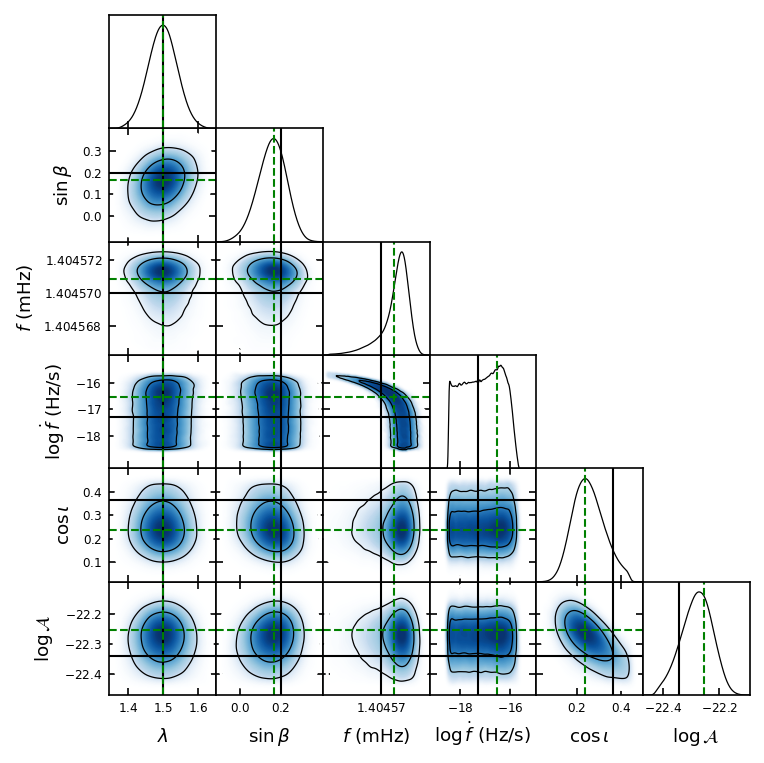}
\caption{\label{fig:low snr} Corner plot of the low SNR signal search. The black lines show the parameters of the injected signal. The blue contours enclose the 68\% and 95\% probability regions of the parameter estimation. The MLE is represented by the dotted green lines. The data, the added signal, and MLE are shown in Figure \ref{fig:low snr signal and mle}.}
\end{figure}

The data of this test is a low SNR signal simulated using FastGB added to the noise realization of LDC1-3v2. The parameters of the signal and the found MLE are listed in Table \ref{tab:parameters low snr}. Due to noise, there exists a parameter set that matches the data slightly better than the true parameter set. Therefore the found MLE has a higher SNR than the actual signal. As a result, the MLE differs slightly from the true signal as shown in Figure \ref{fig:low snr signal and mle}.

The posterior distribution, shown in Figure \ref{fig:low snr}, displays large uncertainties of the found parameters due to the low signal to noise ratio compared to a louder signal shown in Figure \ref{fig:LDC1-3 corner}. The precision of the posterior is limited by the strength of the signal itself. To increase the SNR and therefore the precision, it is beneficial to either decrease the noise by improving the detector or to get stronger signals by increasing the measurement length.

\subsection{Overlapping signals}
\label{section overlap}
LISA will continuously measure signals of estimated $10^7$ GBs in a frequency range of $f \in \left[ 0.1 , 10\right]$ mHz \cite{nelemans2001gravitational, Short-period}. Even though the signals are quasi monochromatic, signals of multiple sources with almost the same frequency will overlap. Therefore we test the pipeline on two overlapping signals, and compare the posterior of the multi signal search with the single signal search.

\begin{table}[!ht]\footnotesize
\caption{Parameters of the two signals to test the pipeline on overlapping signals. The parameters of the two signals only differ at the sky location. The MLE is the maximum likelihood estimate for both of the signals when the data stream contained both signals as pictured in Figure \ref{fig:multisignal_data}.}
\begin{ruledtabular}
\begin{tabular}{ccccc}
Parameter   & Signal 1  & MLE 1     & Signal 2  & MLE 2  \\ \hline
    $\mathcal{A} \times 10^{22}$     &1.36368& 1.497 &1.36368& 1.25\\
    $\beta$ &    $-0.2$   &  $-0.21$   &    0.4    &  0.386\\
  $\lambda$ &      1.4  & 1.404     &      $-1$   & $-1.014$\\
$f$ (mHz)   &  2.01457  & 2.01457003&  2.01457  & 2.01457028\\
 $\log \dot{f}$  &  $-17$    & $-17.47$ & $-17$     & $-19$ \\
 $\iota$    &  0.8      & 0.88      &  0.8      & 0.703 \\
$\phi_0$    &   2       & 1.962     &  2        & 1.813\\
$\psi$      &   1       & 2.552     &   1       & 0.894\\
\end{tabular}
\label{tab:parameters overlap}
\end{ruledtabular}
\end{table}

The naive solution would be to perform a single signal search on the data then subtract the result and conduct another search with the residual as the input of the pipeline. This produces parameter estimations with an error induced by the other signals. This sequential single signal search error can be corrected by doing a global optimization of all found signals using an optimizer with constrained boundaries such as Sequential Least Squares Programming where the MLEs are the initial guess. The original data with the signals of the global solution subtracted is the new input to search for the next signal. This algorithm is repeated until the SNR of the newly found signal is below a threshold $\rho < 8$ or reaches a maximal number of signals which we set to 10. The signals with $\rho < 8$ are not added to the list of the found signals. This algorithm is a similar approach to the slice and dice algorithm \cite{rubbo2006slice}.

Finally, to compute the posterior distributions, all found signals except one get subtracted from the original data and the residual is used to calculate the posterior of the remaining signal.

\begin{figure}[!ht]
\includegraphics[width=0.49\textwidth]{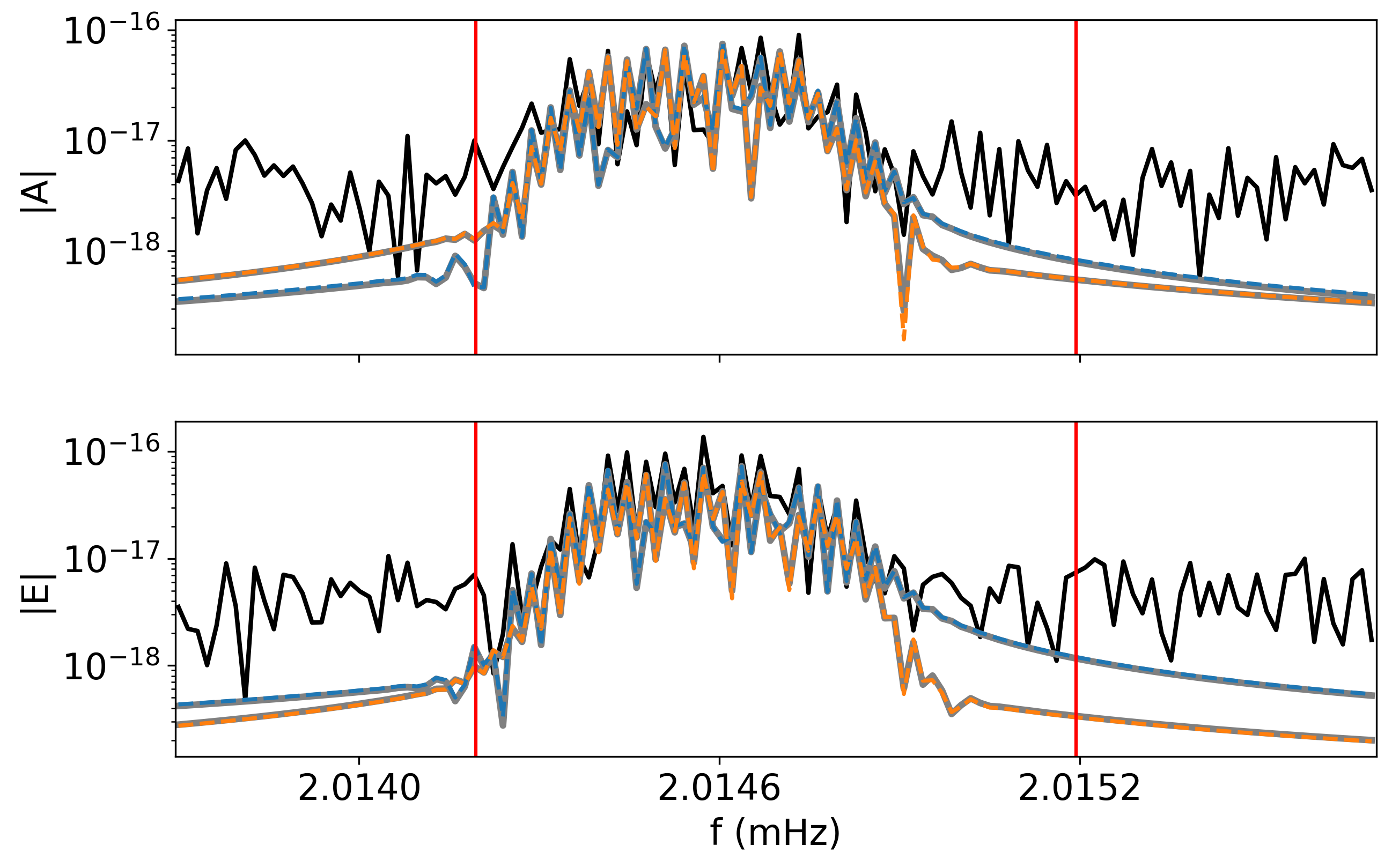}
\caption{\label{fig:multisignal_data} The simulated data strain, shown in black, consists of noise and two overlapping signals. The signals are shown as grey solid lines. The blue dashed line is MLE 1 and the orange dashed line is MLE 2. The vertical lines represent the frequency boundaries within which found signals are accepted. The parameters are listed in Table \ref{tab:parameters overlap}.}
\end{figure}

We test this algorithm on simulated data consisting of instrument noise taken from the LDC1-3v2 and two overlapping signals simulated with FastGB. The signal parameters and found MLEs are listed in Table \ref{tab:parameters overlap}. The LISA response of these signals and the found MLEs are shown in Figure \ref{fig:multisignal_data}. Notice, that both found signals are covering the true signals.

\begin{figure}[!ht]
\includegraphics[width=0.49\textwidth]{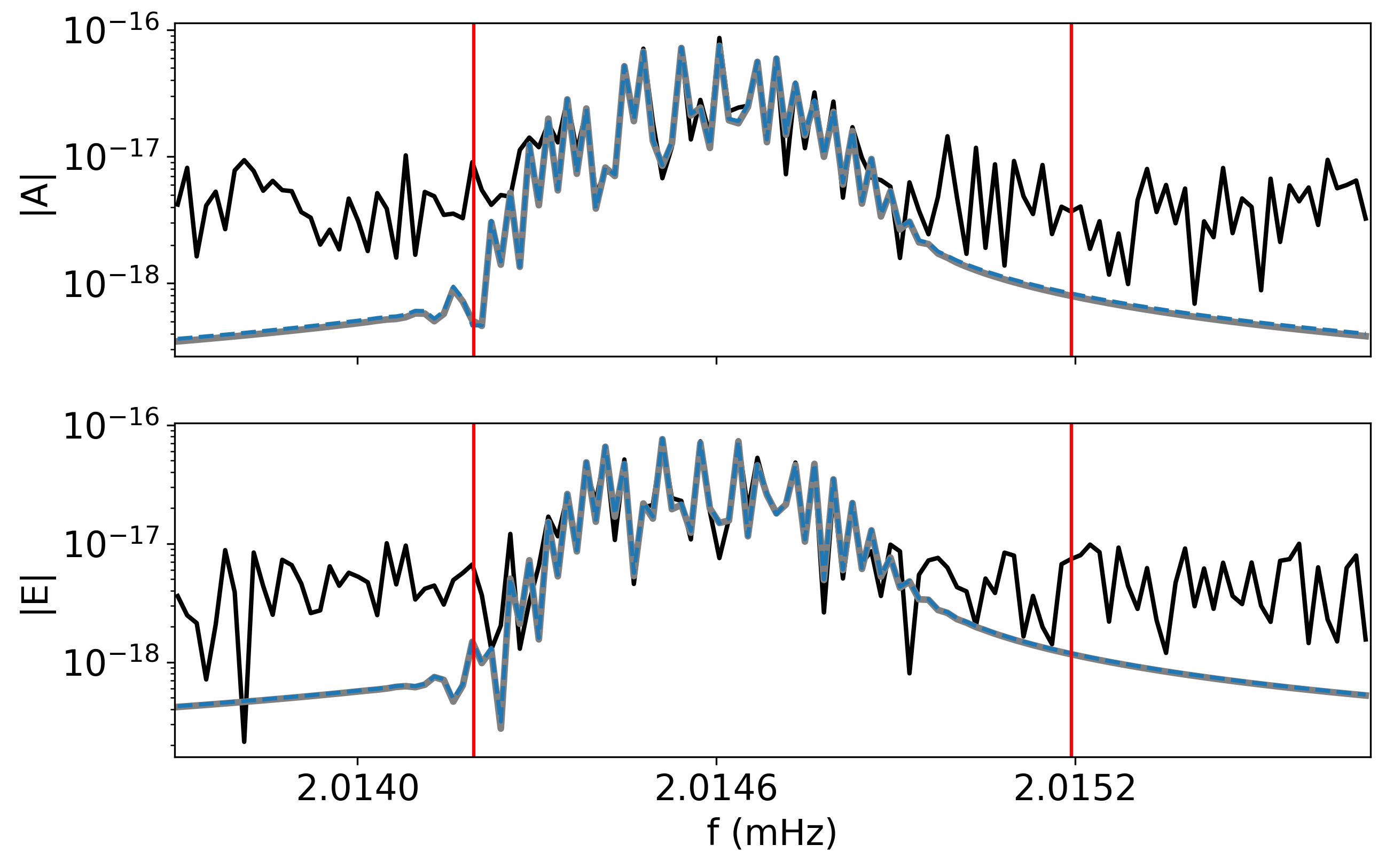}
\caption{\label{fig:singlesignal_data} The simulated data strain, shown in black, consists of noise and one signal. The signal is shown as a grey solid line. The blue dashed line is MLE 1. The vertical lines represent the frequency boundaries within which found signals are accepted. The parameters of signal 1 are listed in Table \ref{tab:parameters overlap}.}
\end{figure}

\begin{figure}[ht!]
\includegraphics[width=0.49\textwidth]{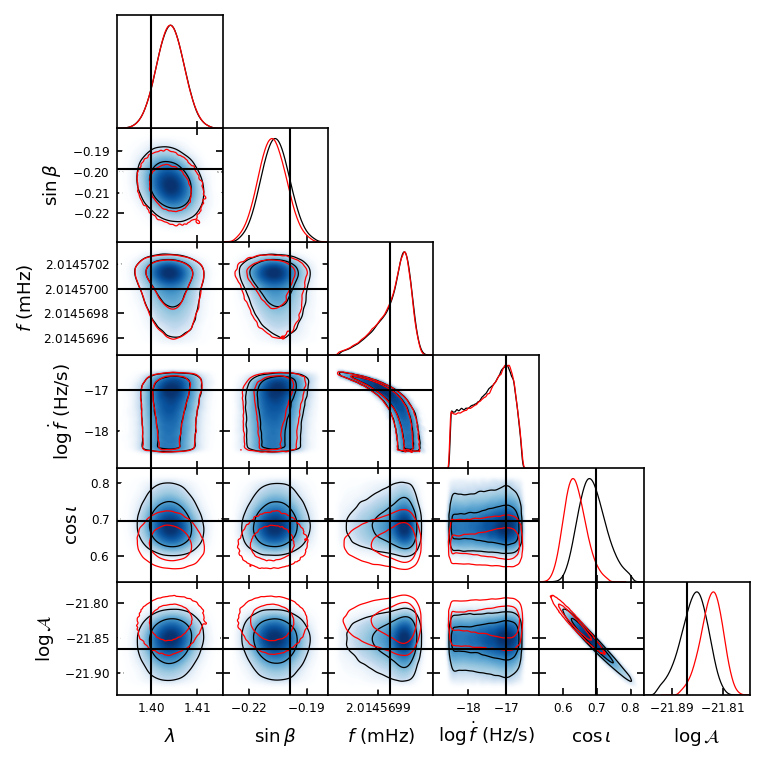}
\caption{\label{fig:multisignal} Posterior distribution of extracting the same signal with two different data streams. The blue shaded contour plot and black histogram show the posterior where the data consists of noise and signal 1 as shown in Figure \ref{fig:singlesignal_data}. The red contour plot and histogram show the posterior of the same signal extracted but the data stream contains additionally signal 2 as shown in Figure \ref{fig:multisignal_data}. The parameters of signal 1 and signal 2 are listed in Table \ref{tab:parameters overlap}. The black lines show the true parameters of signal 1. The contours enclose the 68\% and 95\% probability regions.}
\end{figure}

To check the quality of the solution we compare the posterior of signal 1 of the multi signal data stream with the posterior of the single signal data stream. The single signal data stream is shown in Figure \ref{fig:singlesignal_data}. Figure \ref{fig:multisignal} shows the result of the two searches, where the blue posterior is the result of the single signal search and the red posterior is the result of the multi signal search. The posteriors are almost equivalent for the sky location, frequency, and frequency derivative. For the amplitude and inclination, the posterior differs slightly from each other, nonetheless, both posteriors contain the true value within their 95\% posterior probability regions.

\section{Conclusion}
\label{sec:conclusion}
We present a novel pipeline to accurately extract GB signals from artificial LISA data. In the first step, we identify the MLE of the GB source parameters. In a second step, we model the log-likelihood function enabling a fast sampling to calculate the posterior using the Metropolis-Hastings algorithm. This efficient search enables us to process large sample sizes of 1 million within seconds. Furthermore, using high temperature posteriors as proposal distributions and then sequentially lowering the temperature provides high acceptance rates of 50\% and more. Given the high sample rate and high acceptance rate, employing the Metropolis-Hastings algorithm allows for precise approximations of the posterior distributions.

Multiple tests with increasing difficulty confirm the robustness of our new pipeline. We use the pipeline to quickly extract all signals of the LDC1-3. Furthermore, faint signals of SNR $\rho \approx 8$ are effectively extracted as well. Moreover, the pipeline can be used to globally fit multiple signals, which avoids the systematic error of sequential signal subtraction.

This new pipeline paves the way to efficiently extract a large number of signals from complex simulated LISA data sets such as the LDC1-4 or LDC2a. One core challenge considering the global fit is that some signals are not well confined in one frequency window but overlap with the neighboring window. One way to tackle this issue could be to first subtract signals, found within a frequency window, from the data and then search for signals in the neighboring windows.

\section{Acknowledgements}
We thank the LDC working group for the creation and support of the LDC1-3 and for providing implementations such as FastGB. The authors acknowledge support from the Swiss National Science Foundation (SNF 200021\textunderscore185051).

\bibliography{references}

\end{document}